\documentstyle[aps,prb,multicol]{revtex}

\title{Effect of magnetic impurity correlations on Josephson tunneling}
\author{A.~Bill$^a$, S.A.~Wolf$^b$, Yu.N.~Ovchinnikov$^c$, and V.Z.~Kresin$^a$}
\address{$^a$Lawrence Berkeley Laboratory, University of California,
Berkeley, CA 94720, USA\\
$^b$Naval Research Laboratory, Washington D.C.~20375-5343\\
$^c$L.D.~Landau Institute for Theoretical Physics, Russian Academy of Sciences, Kosygin 2, Moscow, 11733V, Russia}
\begin{document}
\maketitle
{\bf PACS:} 74.50.+r, 74.62Dh \hspace*{0.2cm}

\begin{abstract}
The ordering trend of magnetic impurities at low temperature results
in the frustration of the pair-breaking effect and
induces a ``recovery'' of superconducting properties.
We show that this effect manifests itself in the deviation
of the Josephson current amplitude from the values
obtained within the Ambegaokar-Baratoff and the Abrikosov-Gor'kov
models. We consider both weak and strong-coupling cases. The theory
is applied to describe the experimental data obtained for the
low-$T_c$ superconductor SmRh$_4$B$_4$. We further predict a
``recovery'' effect of the Josephson current in high-temperature
superconductors.
\end{abstract}

\begin{multicols}{2}
\section{INTRODUCTION}

The interaction between mobile charge-carriers
and magnetic impurities is accompanied by a spin-flip process.
In superconductors, this process leads to the pair-breaking
effect\cite{AG,degennes,skalski,OK,kulik,barone1}, that is, to the
destruction of Cooper pairs. As a result,
superconducting properties are strongly affected by the presence of
magnetic impurities. For example, an increase of magnetic impurity
concentration $n_M$ leads to the decrease of the critical temperature
$T_c$ and to gaplesness\cite{AG,degennes,skalski}, to a decrease of the
critical field $H_{c2}$ \cite{OK} or the Josephson current
amplitude $J_c$ \cite{kulik,barone1} (at given temperature).

Most studies including the effect of magnetic impurities
consider the magnetic moments as independent from one another.
Because of the conservation of the total spin (disregarding
spin non-conserving couplings such as the dipole-dipole interaction)
scattering processes involve the spin-flip of both the charge
carrier and the localized magnetic moment.
However, in the region near $T=0$ one should take into account the
correlations between impurity spins.
The ordering trend of the magnetic moments frustrates the spin-flip
process and decreases, therefore, the pair breaking. This decrease
leads to the ``recovery'' effect of
superconductivity. As shown by two of the authors in
Ref.~\onlinecite{OK} the account of
this effect explains the unusual temperature dependence of $H_{c2}$
in high-temperature superconductors and leads to excellent
agreement with all experimental data available so far.

The interplay of superconductivity and ordering of impurity
spins has also been studied in the context of ternary compounds
containing rare earth elements ( $RE$Rh$_4$B$_4$ with $RE=$Sm, Ho,
Er or Lu\cite{vaglio,odoni,ashkenazi}). In some of these materials
($RE=$Sm, Ho, Er) the impurity spin-spin correlations lead to a
magnetic phase transition into an (anti)ferromagnetic state below a
critical temperature $T_m$. SmRh$_4$B$_4$ is of special interest
because it shows coexistence of superconductivity and
antiferromagnetism and displays an unusual temperature dependence of
$H_{c2}(T)$\cite{ashkenazi,ramakrishnan}. Anomalies in the Josephson
critical current have also been observed in this
material.\cite{vaglio}
The experiment shows that the Josephson amplitude is enhanced in the
antiferromagnetic region as compared to the value extrapolated from
the paramagnetic phase.

In the following, we calculate the critical-current amplitude $J_c$
of a Josephson junction in the presence of magnetic impurities and
study the influence of the low-temperature ordering trend.
Our analysis shows that including impurity spin-spin correlations
leads to a sizeable and experimentally observable increase of $J_c$
at low temperatures with respect to the saturated value obtained
when neglecting these correlations.
It would be interesting to carry out experiments verifying our
predictions. The experiment done on SmRh$_4$B$_4$ is also discussed
in the light of our theory.

\section{MAIN EQUATIONS}\label{sec:theory}

\subsection{The Josephson current}\label{subsec:Jc}

Based on the method of thermodynamic Green's functions, one can
obtain a set of equations for the superconducting order parameter
and the renormalization function in the presence of magnetic
impurities:
\begin{eqnarray}
\label{eliMD}
\Delta_nZ_n
= \lambda\pi T\sum_{n'}D_{nn'}\frac{1}{\sqrt{u_{n'}^2 + 1}}
\qquad ,\\
\label{eliMZ}
Z_n = 1 + \lambda\frac{\pi T}{\omega_n}
\sum_{n'}D_{nn'}\frac{u_{n'}}{\sqrt{u_{n'}^2 + 1}}
\qquad ,
\end{eqnarray}
where $D_{nn'} = \Omega^2/
\left[\Omega^2 + (\omega_n - \omega_{n'})^2\right]$ is the phonon
Green's function, $\Omega$ is a characteristic phonon frequency
and $\lambda$ is the coupling constant. Here and in the following we
define $\Delta_n\equiv\Delta({\bf k}_F,\omega_n)$ and
$Z_n\equiv Z({\bf k_F},\omega_n)$ as the order parameter and
renormalization function of the isotropic superconductor
[$\omega_n = (2n+1)\pi T$ with $n=\ldots,-1,0,1,\ldots$; ${\bf k}_F$
is the Fermi wave-vector]. The function $u_n$ that appears in
Eqs.~(\ref{eliMD}) and (\ref{eliMZ}) is solution of the equation
(see Refs.~\onlinecite{AG,degennes,skalski}):
\begin{eqnarray}\label{un}
\frac{\omega_n}{\Delta_n} =
u_n\left( 1 - \frac{\Gamma_s}{\Delta_n}
\frac{1}{\sqrt{u_n^2 + 1}} \right) \qquad .
\end{eqnarray}
$\Gamma_s\sim n_M$ is the spin-flip scattering amplitude ($n_M$ is
the magnetic impurity concentration). In the absence of magnetic
impurities ($\Gamma_s = 0$) the
previous expression reduces to $u_n = \omega_n/\Delta_n$.
Solving Eqs.~(\ref{eliMD})-(\ref{un}) for a given concentration
of impurities (that is, for a given value of the parameter
$\Gamma_s$) and for a given coupling constant $\lambda$ (or $T_c$,
see below) one obtains $u_n$, which allows us to calculate the
Josephson current.

Consider a symmetrical Josephson junction and let us focus on the
stationary Josephson effect. The amplitude $J_c$ of the Josephson
critical current is given by (see, e.g., Ref.~\onlinecite{kulik,barone1})
\begin{equation}\label{Jc}
J_c = \frac{\pi T}{eR} \sum_n \frac{1}{u_n^2 + 1} \qquad ,
\end{equation}
where $R$ is the normal state resistance of the barrier.
This relation applies for both weak and strong-coupling
superconductors. The current depends on the temperature and on the
concentration of magnetic impurities through $u_n$.

In general, the sum over $n$ cannot be evaluated analytically. This
is especially true for the strong-coupling case for which the
dependence of $u_n$, Eq.~(\ref{un}), on $n$ appears not only
through $\omega_n$,
but also through $\Delta_n$ and is thus non-trivial. However, there
are two limiting cases, that are of interest for the present work,
in which an analytical expression can be
derived. Both are within the weak-coupling BCS theory
($\Delta_n \equiv \Delta$). In the absence of magnetic impurities
($\Gamma_s = 0$) one obtains the well-known
Ambegaokar-Baratoff\cite{AB} result:
\begin{equation}\label{JcAB}
J_c^{AB}(T) = \frac{\pi}{2eR} \Delta(T)
\tanh\left(\frac{\Delta(T)}{2k_BT}\right)
\qquad ,
\end{equation}
where $k_B$ is Boltzmann's constant. In the numerical part we
compare this temperature dependence with the one obtained
for a superconductor containing (un)correlated magnetic impurities.
Eq.~(\ref{JcAB}) has two important features. The current saturates as
$T\rightarrow 0$, and is linear near $T_c$. These features
remain when uncorrelated magnetic impurities are added to the
system. In the second limiting case we allow for the presence of
magnetic impurities, but set $T=0$. Then,
\end{multicols}
\begin{eqnarray}\label{Jc0}
eR J_c(T=0) = \left\{ \begin{array}{l@{\quad : \quad}l}
\Delta \left[ \frac{\pi}{2} - \frac{2}{3}\bar{\Gamma} \right] &
\bar{\Gamma} < 1 \\
\Delta\left[ \frac{\pi}{2} - \arctan\sqrt{\bar{\Gamma}^2 - 1}
- \frac{2}{3}\bar{\Gamma}
+ \frac{\sqrt{\bar{\Gamma}^2 -1}}{3\bar{\Gamma}^2}
(2\bar{\Gamma}^2 + 1) \right]
& \bar{\Gamma} > 1
\end{array}\right.
\qquad ,
\end{eqnarray}
\begin{multicols}{2}
with $\bar{\Gamma} = \Gamma_s/\Delta$. This expression is usefull for
the determination of $J_c$ at $T=0$ in the weak-coupling case. For
strong-coupling superconductors one has to use Eqs.~(\ref{eliMD}) and
(\ref{eliMZ}). The numerical solution near $T=0$ can then be
obtained using the Poisson Formula.

\subsection{Spin Ordering of Magnetic Impurities}\label{ssec:SOMI}

The equations of the previous section describe the pair-breaking
effect induced by the time-reversal non-invariant perturbation due
to magnetic impurities.
The theory accounts for such a perturbation through the spin-flip
scattering amplitude $\Gamma_s=1/\tau_s$ (see Eq.~(\ref{un});
$\tau_s$ is the relaxation time).
In the usual picture of independent magnetic moments\cite{AG},
the amplitude $\Gamma_s$ is a temperature independent parameter.
As already stressed in the introduction, the assumption of
independent
impurity spins is a good approximation at high temperatures and low
impurity concentrations, but is not valid in the other cases.
One example where the independent impurity-spin assumption is
not valid is given by systems undergoing a magnetic phase
transition at a critical temperature $T_m$. The observed deviations
from the expected behaviour of superconducting properties
near and below the (anti)ferromagnetic transition temperature of
ternary compounds mentioned in the introduction point towards a
change in the value of $\Gamma_s$ due to spin-spin correlations.
We stress, however, that spin-spin correlations are important even
if one does not observe a magnetic phase transition. For example, the
formation of a spin ``glass'' is related to these correlations.
The importance of magnetic impurity correlations increase with
descreasing temperature.
The anomal behaviour of the critical field $H_{c2}$ observed in the
cuprates can be explained by such correlations\cite{OK}.
Another manifestation of such a ``recovery'' effect is the decrease
in microwave losses \cite{hein} observed under application of an
external magnetic field. In this case, the frustration of the
spin-flip process is due to the external field.

Because of the spin-spin correlations (the ordering trend of magnetic
impurities) the spin-flip scattering amplitude depends on
temperature. This dependence $\Gamma_s(T)$ was studied by
two of the authors in Ref.~\onlinecite{OK} and was applied to the
evaluation of $H_{c2}$. According to this study, $\Gamma_s(T)$
can be represented in the form:
\begin{eqnarray}\label{gamM}
\Gamma_s(T) = \left\{ \begin{array}{l@{\quad : \quad}l}
\Gamma_0 \frac{^{\textstyle 1 + \beta \tau}}{_{\textstyle 1 + \tau}} &
T>T_1\\
\Gamma_0 & T<T_1
\end{array}\right.
\end{eqnarray}
where $\tau = (T - T_1)/\theta$, $T_1 = T_m - \delta T$ with
$\delta T = \alpha\theta$ and $\alpha\simeq 1$. This expression is
valid both for systems that do and do not undergo a magnetic
phase transition. In the former case $T_m$ is the temperature of the
phase transition and $\delta T$ corresponds to the width of the
critical region around $T_m$ (it is such that $T_m -\delta T > 0$).
On the other hand, if there is no phase transition (e.g., when
dealing with the formation of a spin ``glass''), one sets
$T_1= \delta T = T_m = 0$ and $\theta$ is the characteristic
parameter of the ordering. 

The expression (\ref{gamM}) introduced in Ref.~\onlinecite{OK}
describes the ``recovery'' effect in a phenomenological way (a
microscopic treatment will be presented elsewhere).
As can be seen from Eq.~(\ref{gamM}) the spin-flip scattering
amplitude is constant for $T\gg\theta$ (or $T_m$) and
decreases continuously near and below $T\simeq \theta$ (or $T_m$).
Below $T_1$ the scattering amplitude $\Gamma_s$ is a constant
independent of temperature. A
change of only several percents of the value of $\Gamma_s(T)$
at low temperatures can strongly affect the temperature
dependencies of properties such as $H_{c2}$ or $J_c(T)$
(see Ref.~\onlinecite{OK} and below).

Generally speaking, the theory contains three parameters: $\Gamma_0$,
$\beta$ and
$\theta$. To determine their value one has to consider separateley
the case of conventional and high-$T_c$ superconductors.
In the first case, the only existing experiments
studying the effect of ordering on pair-breaking were done on
systems that are (anti)ferromagnetically ordered. A reasonable value
for the characteristic temperature $\theta$ is thus given by the
critical temperature of the phase transition which is of the order
of 1K. Furthermore, the value of
$\Gamma_s(T_c)$ can be obtained within the Abrikosov-Gor'kov theory,
once the experimental value of the critical temperature  $T_c$ in
the presence of magnetic impurities is given. Its value corresponds
to the high-temperature limit of the spin-flip scattering amplitude
(when $\theta \ll T_c$).
We are thus left with one free parameter.

In the case of high-temperature superconductors the analysis of
$H_{c2}$ performed in Ref.~\onlinecite{OK} delivers the values for
all parameters. There is thus no free parameter left for the
high-$T_c$ materials studied in Ref.~\onlinecite{OK}.
For example, for overdoped Tl$_2$Ba$_2$CuO$_6$ studied
experimentally in Ref.~\onlinecite{mackenzie} one has $\Gamma_0 = 95$K,
$\beta = 1.26$, $\theta = 1$K, whereas for Bi$_2$Sr$_2$CuO$_6$
studied in Ref.~\onlinecite{osofski} one has $\Gamma_0 = 105$K,
$\beta = 1.38$ and $\theta = 1.6$K.

Based on Eqs.~(\ref{eliMD})-(\ref{Jc}) and (\ref{gamM}) we can
calculate the amplitude of the Josephson critical current in the
presence of magnetic-impurity ordering at low temperatures.

\section{RESULTS AND DISCUSSION}\label{sec:numres}

We consider the weak and strong coupling cases separately. 
Since, to our knowledge, no experimental study of the ordering
trend on $J_c(T)$ has been undertaken either on conventional or
high-temperature superconductors for systems that do not undergo a
magnetic phase transistion, we chose arbitrary but realistic values
of the parameters and discuss qualitatively what can be expected for
the two types of superconductors.

\subsection{Weak Coupling}\label{ssec:WCN}

For a weak-coupling superconductor $Z\equiv 1$,
$\Delta(i\omega_n) \equiv \Delta$ and one neglects the
dependence on $\omega_{n'}$ of $D_{nn'}$.
Eqs.~(\ref{eliMD})-(\ref{un}) reduce to the Abrikosov-Gor'kov
theory\cite{AG}. For the numerical calculations we determine
$\lambda$ and $\Gamma_s$ from the
experimental values of $T_c$ and $T_{c0}$ (the critical temperature
with and without magnetic impurities). $\lambda$ is obtained from
Eqs.~(\ref{eliMD})-(\ref{un}) for $\Gamma_s = 0$,
i.e.~$\lambda = \ln^{-1}\left( \Omega/T_{c0} \right)$.
On the other hand, $\Gamma_s(T_c)$ can be extracted from the
Abrikosov-Gor'kov
equation for $T_c$ in the presence of magnetic impurities \cite{AG}:
\begin{equation}\label{TcAG}
\ln\left(\frac{T_{c0}}{T_c}\right) =
\psi\left( \frac{1}{2} + \gamma_s \right)
- \psi\left(\frac{1}{2}\right)
\qquad ,
\end{equation}
where $\gamma_s = \Gamma_s(T_c)/2\pi T_c$ is the pair-breaking parameter
and $\psi$ is the psi function.

Fig.~1 shows the temperature dependence of the
Josephson amplitude for different concentrations of magnetic
impurities, that is for different values of $\Gamma_0$. The
temperature and amplitude of the current are normalized to the
critical temperature $T_{c0}$ and the amplitude $J_c^{AB}(0)$
[see Eq.~(\ref{JcAB})] in the
absence of magnetic impurities, respectively. As seen in the figure,
the presence of magnetic
impurities lowers the value of $T_c$ and $J_c(T)$ with respect
to the Ambegaokar-Baratoff result. As stated in the introduction this
is a consequence of the pair-breaking effect. One notes further that
$J_c(T)$ does not saturate as $T\rightarrow 0$
but increase in a temperature range given by $\theta$. The upturn
is the direct signature of the impurity-spin ordering trend, that is,
the temperature dependency of the spin-flip scattering amplitude.
There is a ``recovery'' effect
due to impurity-spin ordering opposing the saturation trend of
$J_c$ for uncorrelated spins. Note that the recovery is not complete
[$J_c(T=0) < J_c^{AB}(0)$] since $\Gamma_s(T\rightarrow 0) \neq 0$.
This is even true for a fully ordered state.
The mechanisms impeding the total recovery of the Ambegaokar-Baratoff
result (as, e.g., dipole-dipole interactions; see also
Ref.~\onlinecite{ovchin}) will be described elswhere.

The ordering trend of local moments leads to a positive
curvature of $J_c(T)$ at low temperatures. Fig.~1 shows that,
the high-temperature linear part becomes dominant as one increases
the concentration of magnetic impurities. The account of correlations
leads thus to a behaviour of $J_c(T)$ that is qualitatively different
from the result obtained for $H_{c2}(T)$\cite{OK}.
Indeed, in the absence of spin correlations, $J_c$ and $H_{c2}$,
considered as functions of temperature, both have a negative
curvature up to $T_c$ and are linear in the vicinity of $T_c$. Adding
magnetic impurities extends the high-temperature
linear part of $J_c(T/T_c)$ to lower temperatures to the expens of
the low-temperature region (that has positive curvature if impurity
correlations are taken into account; see Fig.~1). The
effect on $H_{c2}(T)$ is opposite. For high enough magnetic
impurity concentrations, $H_{c2}(T)$ displays a positive curvature
over the whole temperature range when correlations are taken into
account\cite{OK}.
The concentration of magnetic impurities at which a positive
curvature of $H_{c2}$ was observed experimentally corresponds to 
the case $\Gamma_0 = 8$ of Fig.~1 (i.e.~to the gapless regime).

Fig.~2 gives another representation of the situation presented in
Fig.~1. Compare first the solid and dashed lines. The dashed curve
corresponds to the case $\Gamma_0/T_{c0}=0.6$, $\theta/T_{c0} = 0.1$
and $\beta = 1.3$ (same parameters as the dotted line of figure 1)
whereas the solid line corresponds to the case of uncorrelated spins
and $\Gamma_s/T_{c0} \simeq 0.73$. The latter value was
chosen in such a way that the critical temperature $T_c$ is the same
for both curves. One can see that the result obtained by taking
into account the correlation of impurity spins strongly deviates
from the curve expected when the spins are uncorrelated. This
deviation is already significant at temperatures near $T_c$ for
high enough impurity concentrations, but is
largest near $T=0$. Furthermore, the amplitude of the deviation is
such that it should be experimentally observable.

Another conclusion can be drawn when comparing the dashed and
dotted lines of Fig.~2. These curves were obtained for different
values of the parameter $\beta$ [see Eq.~(\ref{gamM})]. Thus, both
curves include the spin-correlations of the magnetic impurities, but
the strength of the correlations is different. The dashed line
corresponds to $\beta=1.3$ whereas the dotted line is for
$\beta=1.5$. Because $\Gamma_0$ remained unchanged, the two curves
have slightly
different $T_c$'s. The increase of $\beta$, which corresponds to an
increase of the influence of magnetic-impurity correlations, implies
a stronger temperature dependency of $J_c(T)$. Fig.~2 shows that
impurity correlations reduce the high and intermediaire temperature
negative curvature of $J_c(T)$.
Note, finally, that the value at $T=0$ is independent of $\beta$.

It would be interesting to study experimentally the
same high-temperature superconductors as those for which $H_{c2}$
was measured and described in Ref.~\onlinecite{OK}. Indeed, the
values of the parameter for the expression of $\Gamma_s(T)$ are
given by the analysis of $H_{c2}(T)$, leaving no free parameter
(see previous section).
Of special interest would be the case of
YBa$_2$(Cu$_{1-x}$Zn$_x$)$_3$O$_{7-\delta}$ since there are evidences
that the depression of $T_c$ and other superconducting properties
is mainly due to the increase in magnetic impurity concentration
(see also section on strong-coupling superconductors).

Note, finally, that our calculations suggest a possible application
of Josephson junctions based on superconductors doped with magnetic
impurities (as e.g. high-$T_c$ materials). Indeed, because the
temperature dependence of the Josephson current is linear for 
superconductors in the gapless regime (see dash-dotted line of
Fig.~1), the measure of $J_c(T)$ for such systems can be used as a
thermometer\cite{barone2}.

\subsection{Ternary compounds}\label{ssec:ternary}

Up to now we have assumed that the magnetic moments are correlated
but remain disordered. In the case of a magnetic phase transition,
the result differs from above in that the temperature dependency of
$\Gamma_s$ vanishes below $T_m$ [see Eq.~(\ref{gamM})].
This is shown on Fig.~3 for parameters corresponding
to the experimental results on SmRh$_4$B$_4$\cite{vaglio}. The
critical temperature in the absence of magnetic impurities is set
to $T_{c0} = 11.4$K, the value of $T_{c0}$ for LuRh$_4$B$_4$
which has the same structure as the Sm compound, is also
superconducting but has no magnetic moments. The values
$T_c = 1.815$K, $T_m = 0.87$K, $\delta T=0.21$K were taken from the
experiment\cite{vaglio}. The value $\Gamma_0 = 9.75$K was determined
from the measure of $J_c(T_1=T_m - \delta T)$. Finally, the
remaining parameter $\beta = 1.029$ was fixed by the value
$\Gamma_s(T_c)$ chosen in such a way that  $T_c \simeq 1.815$K from
Eqs.~(\ref{gamM}) and (\ref{TcAG}). The values of all parameters have
thus been taken from the experiment and there is no free fitting
parameter left. Regarding the determination of these parameters, one
notes first that we are in the gapless regime
($\bar{\Gamma} = \Gamma_s(T)/\Delta(T) > 1$) and thus in the
situation of the dash-dotted curve of Fig.~1. In addition, the
value of $\beta$ that arises because of the spin-spin correlations
is close to one (value at which the spins are uncorrelated) and thus
much smaller than most of the high-$T_c$ superconductors studied in
Ref.~\onlinecite{OK} (which are of the order $\beta \sim 1.3$).

Comparing our result with the experiment on the ternary compound,
one notes that the behaviour above the temperature $T_1$ (at which
the transition to the antiferromagnetic state is complete) is well
reproduced by our calculation. This demonstrates that the increase
of the current observed near $T_1$ in Ref.~\onlinecite{vaglio} can indeed
be explained by the ordering trend of the magnetic impurities. Below
$T_1$, however, the situation is less clear. If we use
Eq.~(\ref{gamM}) then we observe a distinct discrepancy between
experiment and theory. In the
experiment, the current continues to increase as $T$ is lowered,
whereas the current saturates in our calculation (see dashed line
below $T_1$ in Fig.~3). One reason for this discrepancy lies in the
fact that in metallic systems containing magnetic impurities several
processes occur that were not accounted for in the present analysis.
One of them is connected with symmetry breaking caused by the
impurity\cite{ovchin}. Another reason for the discrepancy is related
to the direct dipole-dipole or RKKY type interaction between the
impurity spins. In the specific case studied in
Ref.~\onlinecite{vaglio}, the occurence of a proximity effect due to
unoxidized parts of the Lu layer of the junction may also be
responsible for the linear behaviour of $J_c(T)$ for $T<T_1$ (see
Ref.~\onlinecite{vaglio}). Further experimental studies should be
carried out on this system to determine which mechanism is the most
relevant to explain the data.

To take into account the additional temperature dependence of the
spin-spin correlations induced by the above processes, we have
considered a temperature dependent $\Gamma_s$ below $T_1$ as well,
but with modified values of the parameters. The experimental
value $J_c(T_1)$ gives $\Gamma_s(T_1)$, $\theta$ is kept unchanged
and $\beta = 1.08$ gives the best fit to the data.
As seen on Fig.~3 (solid line), the temperature dependence of
$\Gamma_s$
below $T_1$ can account well for the experimental result. This is not
a trivial statement because, as shown in the previous figures, the
temperature dependency $J_c(T)$ cannot be modified at ease by
changing the values of the parameter. The fact that one has to
include a temperature dependency of $\Gamma_s$ in the
antiferromagnetic region leads us to the conclusion that not only the
contact interaction but also the other mechanisms (see above and
Ref.~\onlinecite{ovchin}) are important in this material.

\subsection{Strong Coupling}\label{ssec:SCN}

We study the influence of strong coupling onto the results of the
last section. This case is numerically more difficult to handle, due
to the fact that the renormalization function has to be considered
and that all three unknown functions $\Delta_n$, $Z_n$ and $u_n$
depend on $n$. As before, we determine $\Gamma_s$ and $\lambda$ from
the equations
for $T_c$ and $T_{c0}$. For $T_c$ one linearizes  Eq.~(\ref{eliMD}).
As for $T_{c0}$ ($\Gamma_s = 0$) we use the expression derived
by one of the authors\cite{kresin}:
\begin{equation}\label{Tc0S}
T_{c0}
= \frac{0.25\Omega}{\left(e^{2/\lambda} - 1\right)^{1/2}}\qquad,
\end{equation}
where $\Omega$ is the characteristic phonon frequency.
Fig.~4 shows a generic case for the strong coupling limit. There is
no qualitative difference between the weak and strong coupling
results. The parameters in Fig.~4 were chosen so as to correspond to
the case of YBa$_2$(Cu$_{1-x}$Zn$_x$)$_3$O$_{7-\delta}$ with
$x\simeq 0.04$ ($T_c \simeq 55$K)\cite{zagoulev}$^a$. This result
leads us to conclude that the overall qualitative behaviour obtained
for weak-coupling superconductors is not modified by increasing the
strength of the coupling. In particular, one expects a nearly linear
behaviour of $J_c(T)$ for high-$T_c$ materials where $T_c$ has been
strongly reduced by the presence of magnetic impurities and are in
the gapless regime. Examples of such systems are Zn-doped
YBa$_2$Cu$_3$O$_{7-\delta}$\cite{zagoulev}$^a$ and
La$_{1.85}$Sr$_{0.15}$CuO$_4$\cite{zagoulev}$^{b,c}$, as well as
other under- and overdoped materials studied in Ref.~\onlinecite{OK}
(Tl$_2$Ba$_2$CuO$_6$, Bi$_2$Sr$_2$CuO$_6$, or
Sm$_{1.85}$Ce$_{0.15}$CuO$_{4-y}$). It would be interesting to
perform measurements on these (or other) systems to verify our
predictions. As stated earlier, the above mentioned materials
are of special interest, since all parameters necessary
to calculate $\Gamma_s(T)$ and $J_c(T)$ [Eqs.~(\ref{gamM}) and
(\ref{Jc}), respectively] were determined from the study of
H$_{c2}$\cite{OK}.

\section{CONCLUSION}\label{sec:concl}

In summary, we have calculated the amplitude of the Josephson
current $J_c(T)$ in the presence of magnetic impurities taking into
account their ordering trend at low temperatures. We have shown that
the ``recovery'' effect introduced in Ref.~\onlinecite{OK} is also occuring
for the Josephson current $J_c(T)$. The effect manifests itself in
an upturn of $J_c(T)$ at low temperatures (instead of the saturation
occuring for uncorrelated magnetic moments). The effect is sizeable
and qualitatively similar in weak and
strong-coupling superconductors. The theory was used to explain the
experimental results obtained for SmRh$_4$B$_4$ and leads to good
agreement for temperatures above the magnetic phase transition. We
have also presented results for high-$T_c$ materials.

It would be interesting to measure the effect in both conventional
and high-temperature superconductors so as to verify our predictions.
The effect might be larger in the latter (for overdoped or Zn-doped
systems) as indicated by the unusal temperature dependence of
$H_{c2}$\cite{OK}, since most of these materials are in the gapless
regime.\\

A.B.~is grateful to the Swiss National Science Foundation and the
Naval Research Laboratory for the support. The work of V.Z.K.~was
supported by the U.S.~Office of Naval Research Contract under
No.~N00014-96-F0006.

\end{multicols}
\newpage

{\bf Figure captions:}\\[0.5cm]

Fig.~1: Temperature dependence of the Josephson current amplitude
$J_c(T)$ for different magnetic impurity concentrations. The
temperature and the current amplitude are normalized to the values
$T_{c0}$ and $J_c^{AB}(T=0)$ in the absence of magnetic impurities,
respectively. $\beta=1.3$, $\theta/T_{c0} = 0.1$ and 
$\Gamma_0/T_{c0} = 0$ (Ambegaokar-Baratoff; solid line),
$\Gamma_0/T_{c0} = 0.4$ (dashed), $\Gamma_0/T_{c0} = 0.6$ (dotted)
$\Gamma_0/T_{c0} = 0.8$ (dotted-dashed).\\[0.5cm]
Fig.~2: Temperature dependence of the Josephson current for different
values of $\beta$. Solid line: $\beta = 1$, dashed line: $\beta =
1.3$, dotted line $\beta = 1.5$. $\Gamma_0/T_{c0} = 0.6$, $\theta/T_{c0} = 0.1$.\\[0.5cm]
Fig.~3: Josephson current for SmRh$_4$B$_4$. Solid line: theoretical
result with $\Gamma_s(T)$, dashed line: $J_c(T)$ for
$\Gamma_s(T<T_1)=\Gamma_0$, dots: experimental points from
Ref.~\onlinecite{vaglio}. The parameters are given in the
text.\\[0.5cm]
Fig.~4: Influence of impurity-spin correlations on the Josephson
current amplitude for strong-coupling superconductors. solid line:
uncorrelated spins, dashed line: correlated impurity spins. The
parameters correspond to the case of
YBa$_2$(Cu$_{1-x}$Zn$_x$)$_3$O$_{7-\delta}$ with $x\simeq 0.04$ (see
text).

\newpage
\unitlength1cm
\begin{figure}[h]
\begin{center}
\input{Fig1.ps}
\end{center}
\vspace*{2cm}
\caption{}
\end{figure}
\vspace*{5cm}
\centerline{A.~Bill et al.}

\newpage

\unitlength1cm
\begin{figure}[h]
\begin{center}
\input{Fig2.ps}
\end{center}
\vspace*{2cm}
\caption{}
\end{figure}
\vspace*{5cm}
\centerline{A.~Bill et al.}

\newpage

\unitlength1cm
\begin{figure}[h]
\begin{center}
\input{Fig3.ps}
\end{center}
\vspace*{2cm}
\caption{}
\end{figure}
\vspace*{5cm}
\centerline{A.~Bill et al.}

\newpage

\unitlength1cm
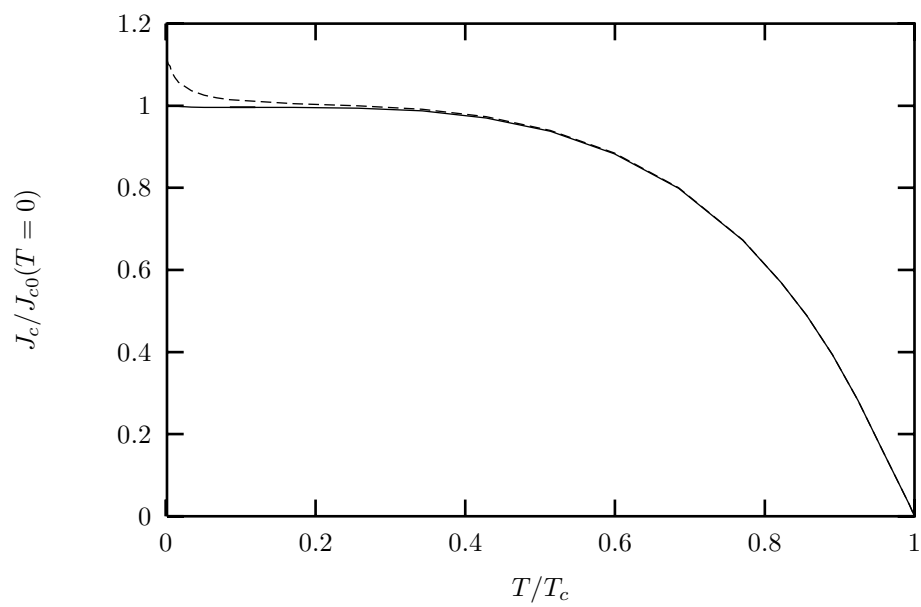
\begin{figure}[h]
\begin{center}
\input{Fig4.ps}
\end{center}
\vspace*{2cm}
\caption{}
\end{figure}
\vspace*{5cm}
\centerline{A.~Bill et al.}

\end{document}